\documentstyle[preprint,aps,epsfig]{revtex}
\begin{document}

\tighten
\draft
\preprint{
\vbox{
\hbox{\today}
\hbox{Tashkent}
}}

\newcommand{\re}[1]{(\ref{#1})}
\newcommand{\lab}[1]{\label{#1}}
\newcommand{\ci}[1]{\cite{#1}}
\renewcommand{\baselinestretch}{1.25}
\newcommand{\bfr}{\begin{flushright}}
\newcommand{\bfl}{\begin{flushleft}}
\newcommand{\efl}{\end{flushleft}}
\newcommand{\efr}{\end{flushright}}
\newcommand{\bc}{\begin{center}}
\newcommand{\ec}{\end{center}}
\newcommand{\be}{\begin{equation}}
\newcommand{\ee}{\end{equation}}
\newcommand{\bea}{\begin{eqnarray}}
\newcommand{\eea}{\end{eqnarray}}
\newcommand{\ba}{\begin{array}}
\newcommand{\ea}{\end{array}}
\newcommand{\edc}{\end{document}}
\newcommand{\ul}{\underline}
\newcommand{\ri}{\rightarrow\infty}
\newcommand{\li}{\leftarrow\infty}
\newcommand{\ra}{\rightarrow}
\newcommand{\la}{\leftarrow}
\newcommand{\ds}{\displaystyle}
\newcommand{\dsf}{\displaystyle\frac}
\newcommand{\dt}{\Delta{t}}
\newcommand{\il}{\int\limits}
\newcommand{\pal}{\partial}
\newcommand{\xxx}{{\it{X}}}
\newcommand{\bone}{{\bf 1}}
\newcommand{\gComment}[1]{}

\title{
Approximate ground state of a confined Coulomb anyon gas in an external
magnetic field}


\author{B. Abdullaev$^1$, G. Ortiz$^2$, U. R\"{o}ssler$^{3}$,
M. Musakhanov$^1$, and A. Nakamura$^{4}$ }

\address{$^1$
 Theoretical Physics Dept., Uzbekistan National University,
 Tashkent 700174, Uzbekistan}
\address{$^2$
Theoretical Division, Los Alamos National Laboratory,
Los Alamos, NM 87545, USA}
\address{$^3$
Institute for Theoretical Physics, University of Regensburg, D-93040
Regensburg, Germany}
\address{$^4$
RIISE, Hiroshima University, Japan}

\date{Received \today }

\maketitle

\begin{abstract}
We derive an analytic, albeit approximate, expression for the ground
state energy of $N$ Coulomb interacting anyons with fractional
statistics $\nu$, $0\leq|\nu|\leq1$, confined in a two-dimensional well
(with characteristic frequency $\omega_0$) and subjected to an external
magnetic field (with cyclotron frequency $\omega_c$). We apply a
variational principle combined with a regularization procedure which
consists of fitting a cut-off parameter to existing exact analytical
results in the non-interacting case, and to numerical calculations for
electrons in quantum dots in the interacting case. The resulting
expression depends upon parameters of the system $|\nu|, N, \omega_0,
r_0, a_B$ and $\omega_c$, where $r_0$ represents a characteristic unit
length and $a_B$ the Bohr radius. Validity of the result is critically
assessed by comparison with exact, approximate, and numerical results
from the literature.

\end{abstract}



\newpage

\section{Introduction}
The exchange statistics of particles whose orbital motion is restricted
to two space dimensions (2$D$) differs substantially from the 3$D$
case. The topology of their (multiply connected) configuration space
allows for fractional statistics  \ci{lei}, characterized by a
continuous parameter $\nu$ which labels the possible one-dimensional
representations of the braid group. For particles in 2$D$, $\nu$ may
attain values between 0 (for bosons) and 1 (for fermions), thus 2$D$
particles are called anyons \ci{wil2}. The concept of anyons has been
used to describe quasi-particle excitations in the fractional quantum
Hall regime \ci{fis,ler,kh} and in high-$T_c$ superconductors
\ci{wil1}.

A system of particular current interest is that of 2$D$ electrons in a
parabolic confinement potential, the so-called quantum dot or
artificial atom \ci{hawrylak}. These systems, realized in semiconductor
nanostructures, are objects of fundamental studies of ground state
properties of interacting $N$-particle systems and have also a
potential for applications in quantum information and computation
\ci{loss}. Exact closed-form solutions of the problem is reduced to a
few simple cases due to its intrinsic mathematical complexity
\ci{t,jq}. Typically, ground state calculations make use of numerical
simulations for individual choices of the parameters of the system. It
would be desirable to have an accurate, albeit approximate, analytical
expression describing the ground state energy of such a system as a
function of the parameters of the system (including an external
magnetic field applied perpendicular to the plane of the dot). In order
to derive such a formula we make use of the anyon concept including the
effect of the Coulomb interaction.

Anyons in a parabolic confining potential (with and without an external
magnetic field) has been the subject of several investigations in the
past. For the $N=2$ case an exact solution to the spectral problem
exists \ci{lei,wil2,fis,kh}. For the noninteracting case, its
generalization to $N=3$ (without magnetic field) is considered in
\ci{wu}. Numerical calculations have been performed for the lower part
of the energy spectrum for $N=3$ \ci{spor1m} and $N=4$ \ci{spor2}.
Making use of the separation of the center-of-mass and relative motion
one finds in the two-anyon case that the ground state energy for the
relative motion is a linear function of $\nu$ for $0\leq\nu\leq 1$. For
$N=3$ and $4$ \ci{spor1m,spor2,kh} the ground state energy of the
relative motion is a linear function of $\nu$ only near the bosonic
limit $\nu\simeq 0$. The ground state in the fermionic end ($\nu=1$)
is continuously connected to an excited state of the bosonic spectrum
($\nu=0$) and consequently - when adding the lowest energy of the
center-of-mass motion (which does not depend on $\nu$) - one finds for
$N=3,4$ (and likely for all higher $N$ \ci{chitra}) a nonlinear $\nu$
dependence of the ground state energy. In the case of interacting
anyons in no external magnetic field the energy spectrum for two anyons
at some fixed values of the Coulomb interaction parameter was found
analytically in \ci{v} and approximately, for two and three anyons, in
\ci{mgg}.

In the presence of an external magnetic field one can study the
interplay between the statistical and the physical magnetic flux. This
has been done analytically for the $N$=2 noninteracting anyon case with
confinement \ci{johcom}, while the cases with $N>2$ have been
investigated preferentially without confinement \ci{anmf,vo,il}. The
ground state for the $N=3$ case including confinement and magnetic
field was calculated in Ref.~\ci{kmo}. The case with the applied
external magnetic field and Coulomb interaction for two anyons in a
harmonic potential was considered in Ref. \ci{roy} and for two and
three anyons in Ref. \ci{per}. Ref.~\ci{kh} provides a review about all
these studies.

In our treatment of the $N$-anyon problem we make use of the bosonic
representation of anyons that works with a gauge vector potential to
account for the fractional exchange statistics but allows to use a
product ansatz for the $N$-body wave function. We apply a variational
principle by constructing this wave function from single-particle
gaussians of variable shape. It is well-known from perturbative ground
state calculations for anyons in an oscillator potential that the
expression for the ground state energy has a logarithmic divergence
connected with a cut-off parameter for the interparticle distance
\ci{cho,ouvgroup,ac1,sen,ac2,chitra}. We face the same problem in our
variational treatment. Making use of the physical argument (see
Ref.~\ci{ler}) that for $\nu\neq 0$ this distance has to have some
finite value, we regularize the formula obtained for the ground state
energy by an appropriate procedure that takes into account some
existing exact analytical results, in the case without Coulomb
interaction, and numerical results for electrons in quantum dots, in
the case with Coulomb interaction. Our formula, which is an approximate
closed-form expression depending upon $\nu ,N$, $\omega_0$ (confinement
parameter), $r_0/a_B$ (Coulomb interaction parameter), where
$r_0=(\hbar/(M\omega_0))^{1/2}$ and $a_B=\hbar^2/(Me^2)$ (in the
presence of a magnetic field also of the parameter $\omega_c/\omega_0$,
and we need to replace $\nu$ by  $|\nu|$), will be compared to exact,
approximate and  numerical results for quantum dots reported in the
literature.

The paper is organized as follows: In Section II we describe the system
and motivate the ansatz for the variational treatment, in Section III
we present the calculations without, and in Section IV with a
homogeneous magnetic field for the case without Coulomb interaction.
Calculations including Coulomb interactions are presented in Sections
V and VI. Finally, Section VII summarizes the main conclusions.


We would like to note the existence of two seemingly unrelated notions
of anyonic statistics in the literature. One originally introduced  in
first quantization in the coordinate representation, and another
derived within the framework of quantum field theory. In both cases the
original motivation to introduce such particles was basically as an
inherent possibility in the kinematics of (2+1)-dimensional quantum
mechanics and clearly the concepts, if correctly implemented, should be
equivalent whether one uses first quantization in the coordinate
representation or second quantization \cite{ouralg,oursanyon}. Within
the framework of quantum field theory fermions can be kinematically
transformed into hard-core bosons (through statistical transmutation)
but not into canonical ones, thus preserving the exclusion statistics
properties of the particles. More generically, the Hamiltonian spectra
of particles sharing the same exclusion statistics can be connected
through a continuous mapping.  The anyon notion used in the present
manuscript is consistent with the one developed in the framework  of
the bosonic representation in first quantization.
Had we used the fermionic representation we would have ended up in an
excited bosonic state.

\section{Interacting Anyons in a $2D$ parabolic well in the presence of
an external magnetic field: General setup}

The Hamiltonian of $N$ spinless anyons of mass $M$ and charge $e$
confined to a $2D$ parabolic well, interacting through Coulomb
repulsions, and in the presence of an external homogeneous magnetic
field, $\vec{H}=H\vec{e}_z$ ($\omega_c=|eH|/(Mc)$), is given by
\be
\hat H=\dsf{1}{2M}\ds\sum_{k=1}^N\left[\left(\vec p_k-(\vec
A_{\nu}(\vec r_k)+ e\vec{A}_{ext}(\vec r_k)/c)\right)^2+M^2\omega_0^2
|\vec{r_k}|^2\right] +\dsf{1}{2}\ds\sum_{k,j\not=k}^N\dsf{e^2}{|\vec
r_{kj}|}.
\lab{gsetup1}
\ee
Here $\vec r_k$ and $\vec p_k$ represent the position and momentum
operators of the $k$th anyon in two space dimensions,
\be
\vec A_{\nu }(\vec r_k)=\hbar\nu\ds\sum_{j\not=k}^N\dsf{\vec e_z
\times\vec r_{kj}} {|\vec r_{kj}|^2}
\lab{gsetup2}
\ee
is the anyon gauge vector potential \ci{wu,lau}, $\vec r_{kj}=\vec
r_k-\vec r_j$, and $\vec e_z$ is the unit vector normal to the 2$D$
plane. The factor $\nu$ determines the fractional statistics (or spin)
of the anyon: it varies between $\nu=0$ (bosons) and $\nu=1$
(fermions).The external magnetic field enters by minimally coupling
the vector potential  $\vec{A}_{ext}(\vec{r}_k)=\vec
H\times\vec{r}_k/2$.

In order to find an analytic expression for the ground state energy
as a function of  $\nu ,N$, $\omega_0$, $r_0/a_B$ and
$\omega_c/\omega_0$ (in the presence of a magnetic field $\nu$  is
replaced by  $|\nu|$ (see Section IV below)) we employ a variational
scheme by minimizing the expression for the total energy
\be
E=\dsf{\int \Psi_T^*(\vec R)\hat H \Psi_T(\vec R) \ d\vec R}{\int
\Psi_T^*(\vec R) \Psi_T(\vec R) \ d\vec R} \ ,
\lab{gsetup3}
\ee
with a trial wave function $\Psi_T(\vec R)$ depending on the
configuration $\vec R = \{\vec r_1....\vec r_N\}$ of the $N$ anyons. To
motivate the choice of $\Psi_T(\vec R)$ we invoke the mean-field
approximation to the gauge vector field
\be
\overline {\vec{A}_{\nu}}(\vec r) =
\dsf{1}{2}\vec B_\nu\times\vec r
\lab{gsetup4}
\ee
introduced by Fetter, Hanna, and Laughlin \ci{fet}. This
single-particle vector potential can be understood as that of a
homogeneous ``magnetic'' field $\vec B_\nu = 2\pi\rho\hbar\nu\vec e_z$
connected with the carrier density $\rho$ and the anyonic factor $\nu$
(note: $\vec B_\nu$ vanishes in the bosonic limit). By analogy to a
physical magnetic field one can introduce a ``magnetic'' length
$l_{\nu} = (\hbar/B_\nu)^{1/2}$. The other characteristic length of the
system is the mean distance between particles $r_0=1/\sqrt{\pi\rho}$.
Taking into account only this mean gauge vector field (and not the
external parabolic confining potential) one obtains a Landau spectrum
\ci{gal} and it is reasonable, in the bosonic representation of anyons
when the many-body wave function takes the product form
\be
\Psi_T (\vec R)=\prod_{k=1}^N\psi_T (\vec r_k)\:,
\lab{gsetup5}
\ee
to adopt the single-particle trial functions $\psi_T(\vec r_k)$ in the
form
\be
\psi_T(\vec
r_k)=C\exp\left(-(\alpha'+\nu)\dsf{(x_k^2+y_k^2)}{2r_0^2}\right)
\lab{gsetup6}
\ee
typical for the lowest Landau level. Here $C$ is a normalization
constant and $\alpha'$ a variational parameter. To include the
external confining potential we identify $r_0$ with the
characteristic length $(\hbar/M\omega_0)^{1/2}$ of this harmonic
oscillator.
When energies are expressed in units of $\hbar\omega_0$ and
lengths in units of $r_0$ the normalized trial wave function reads
\be
\Psi_T(\vec R)=\left(\dsf{\alpha}{\pi}\right)^{N/2}\prod_{k=1}^N
\exp\left(-\alpha\dsf{(x_k^2+y_k^2)}{2}\right),
\lab{gsetup7}
\ee
where $\alpha=\alpha'+\nu$.

In evaluating the expectation value $E$ (Eq.~\re{gsetup3}) it is
convenient to consider the local energy  $E_L(\vec R)=\Psi_T^{-1}(\vec
R)\hat H \Psi_T(\vec R)$ \ci{ceperley}. In general $E_L(\vec R)$ is a
complex function
\be
E_L(\vec R)=Re E_L(\vec R)+i Im E_L(\vec R)
\lab{gsetup8}
\ee
with
\be
Im E_L(\vec R) = - \alpha\ds\sum_{k=1}^N((\vec A_{\nu}(\vec
r_k)+ e\vec{A}_{ext}(\vec r_k)/c)\cdot\vec r_k).
\lab{gsetup9}
\ee
However, evaluation of the expectation value $E=\int \Psi_T(\vec R)\
E_L(\vec R)\Psi_T(\vec R) \ d\vec R$ immediately yields
\be
\int \Psi_T(\vec R)\ Im E_L(\vec R)\Psi_T(\vec R) \ d\vec R=0,
\lab{gsetup10}
\ee
and, therefore, the only quantity to consider in the following is $Re
E_L(\vec R)$. Before proceeding, we would like to emphasize that the
absolute ground state of the anyon system is a non-analytic function
of $\nu$. Our calculations will simply provide a smooth interpolation.

\section{Non-interacting case and $H=0$}

In the non-interacting case, in the absence of an external magnetic
field, the local energy is
\be
Re E_L(\vec R)=\ds\sum_{k=1}^N[\alpha+\dsf{x_k^2+y_k^2}{2}(1-
{\alpha}^2)+\dsf{\nu^2}{2}(\vec A_{\nu}(\vec
r_k))^2] .
\lab{ncnh1}
\ee
The expectation value of $Re E_L(\vec R)$ can be easily calculated for
the first two terms of Eq. \re{ncnh1}. The last term contributes with
integrals of the form $\int \Psi_T(\vec R)\ \dsf{\vec r_{kj} \cdot \vec
r_{kl}} {|\vec r_{kj}|^2 |\vec r_{kl}|^2} \Psi_T(\vec R) \ d\vec R$,
which fall into one class of $N(N-1)$ integrals with $j=l$ and a second
class of $N(N-1)(N-2)$ integrals with $k\neq j, k\neq l$ and $j\neq l$.
The first class of integrals can be evaluated using \ci{gr}
\be
\int_{0}^{\infty} Ei(ax)e^{-\mu x}dx=-\dsf{1}{\mu}\ln
\left(\dsf{\mu}{a}-1\right)
\lab{ncnh2}
\ee
with $a>0$, $Re \mu>0$ and $\mu>a$. ($Ei(y)= - \int_{-y}^{\infty} e^{-
z}dz/z$ is the exponential integral with $y<0$.) The result is
\be
\int \Psi_T(\vec R) \dsf{1}{|\vec r_{kj}|^2}\Psi_T(\vec R) \ d\vec
R\approx \alpha\ln\left(\dsf{1}{2 \delta}\right) \ ,
\lab{ncnh4}
\ee
which displays a logarithmic divergence with the cut-off parameter
$\delta$ tending to zero. The integrals of the second class yield
\be
\int \Psi_T(\vec R)\ \dsf{\vec r_{kj} \cdot \vec r_{kl}} {|\vec
r_{kj}|^2 |\vec r_{kl}|^2} \Psi_T(\vec R) \ d\vec R=-\alpha G \ ,
\lab{ncnh3}
\ee
where $G=3^{1/2}\ln(4/3)$. Putting together all these different
contributions one obtains
\be
E=\frac{N}{2}\left( {\cal N} \ \alpha + \frac{1}{\alpha}\right) \
\lab{ncnh5}
\ee
with
\be
{\cal N}=1+\nu^2 (N-1)[\ln\left(\dsf{1}{2 \delta}\right)- G(N-2)] \ ,
\ee
which attains a minimum ($\dsf{dE}{d\alpha}=0$) for
\be
\alpha_0={\cal N}^{-1/2} .
\lab{ncnh6}
\ee
Thus, the resulting expression for the ground state energy is
\be
E_0=N \ {\cal N}^{1/2} .
\lab{ncnh7}
\ee

The logarithmic divergence displayed in $E_0$ when $\delta\rightarrow 0$ has
also been found in other approximate perturbative treatments of the problem
and is widely discussed in the literature. To remedy this problem, various
solutions were introduced: In Ref.~\ci{cho} a hard-core centrifugal term and
in Ref.~\ci{ouvgroup} a pair correlation term were introduced in the trial
wave function, while in Ref.~\ci{ac1} both modifications of $\psi_T(\vec r)$
were used. An artificial repulsive delta-like potential was assumed in
Refs.~\ci{sen,chitra,ac2} when the unperturbated ground state wave function
is a product of single particle (gaussian) wave functions. Here we
assume as in Ref.~\ci{ler} that the cut-off parameter $\delta$ cannot
be zero for $\nu>0$, away from the bosonic limit, since it corresponds
to the square of the nearest distance between the particles. Thus, for
anyons in the parabolic confining potential $\delta$ is definitely
smaller than 1 (in units of $r_0^2$). In the following we determine
$\delta$ by fitting to appropriate results for special values of the
parameters of the system.

\begin{figure}
\begin{center}
\includegraphics[angle=0,width=10.5cm,scale=1.0]{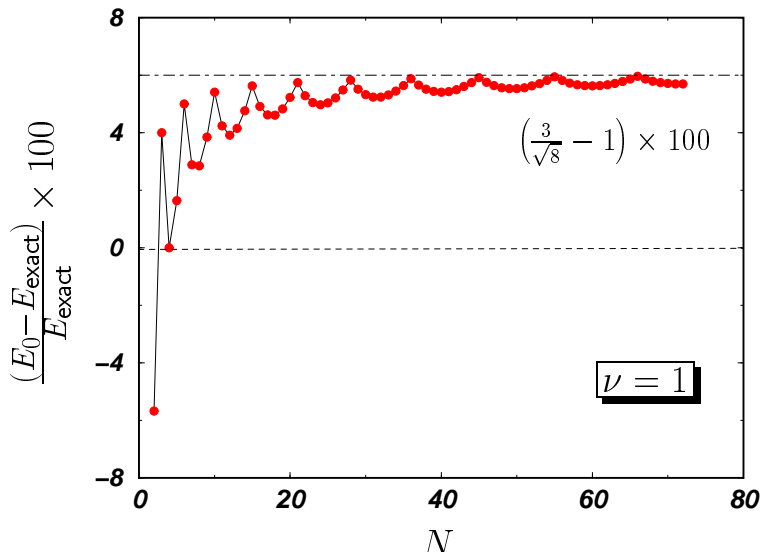}
\label{fig1}
\caption{
Relative deviation (in percent) of the approximate ground state energy
$E_0$, Eq. \protect \re{ncnh10}, from the exact ground state energy,
$E_{\sf exact}$, for up to $N$=72 noninteracting fermions ($\nu=1$) in
a parabolic confining potential. The dashed-dotted line indicates the
asymptotic ($N\rightarrow\infty$) value.
}
\end{center}
\end{figure}

Wu \ci{wu} has computed the ground state energy of $N$ anyons in a 2$D$
parabolic potential near the bosonic limit $\nu\simeq 0$ and obtained
\be
E \approx [N+N(N-1)\nu/2].
\lab{ncnh8}
\ee
To regularize the expression for $E_0$ we make use of this result by
expanding $E_0$, Eq.~\re{ncnh7}, for $\nu \rightarrow 0$ and
identify the leading term in $\nu^2$ with the term linear in $\nu$ of
Eq.~\re{ncnh8}, with the result
\be
\delta= \dsf{1}{2} \exp\left[ -\dsf{1+\nu G(N-2)}{\nu}\right].
\lab{ncnh9}
\ee
With this value of the cut-off parameter the final analytic expression
for the ground state energy is (see also Ref. \ci{amn})
\be
E_0=N[1+\nu(N-1)]^{1/2}.
\lab{ncnh10}
\ee

By construction, it is evident that this formula reproduces the result
of Wu \ci{wu} in the bosonic limit $\nu\rightarrow 0$. Less trivial,
however, is the asymptotics in the fermionic end: For large $N$ it is
consistent (up to a numerical factor) with the approximate expression
$E \approx \nu ^{1/2}N^{3/2}$ of Chitra and Sen \ci{chitra} calculated
perturbatively from the bosonic end for $\nu>1/N$. These authors have
also studied the fermionic end $\nu\simeq 1$ and found for $N\gg 1$ the
expression $E\approx (8N^3)^{1/2}/3$. This formula is asymptotic to the
exact ground state energy $E=N_{cl}(1+8N_{cl})^{1/2}/3$ for $N_{cl}$
fermions filling the first $K$ closed shells \ci{kh},  where
$N_{cl}=K(K+1)/2$.
Note that Eq.~\re{ncnh10} provides a monotonically increasing function of
$\nu$ while in the closed-shell case the exact fermionic end has lower
energy (by a factor $8^{1/2}/3$) than the one calculated from the bosonic end.

In Fig. ~1 we compare exact ground state energies (the sum of occupied
harmonic oscillator states), for up to $N=72$ fermions ($\nu=1$), with
the results obtained from Eq.~\re{ncnh10}. As it turns out, the
relative deviation does not exceed $6\%$. Fig. 2 shows the relative
deviation for 2,3 and 4 anyons as a function of $\nu$ (the exact ground
state energies here are taken from Refs. \ci{lei,wil2,spor1m,spor2}).
In this figure we have considered all the cases for which the exact
ground state energies are known. When the number of anyons $N$
increases, the absolute ground state of the system is a non-analytic
function of $\nu$ because there is approximately $N^{1/2}$ number of
level crossings \ci{chitra}. Since our formula for $E_0$,
Eq.~\re{ncnh10}, has the same ($N\rightarrow \infty$) asymptotics as
the formula obtained by Chitra and Sen \ci{chitra} one expects that,
for $0<\nu<1$, this relative deviation will be bounded as the number of
particles is increased.

It should be noted, that due to our regularization procedure the ground
state energy obtained, Eq.~\re{ncnh10}, is not an upper bound to the
ground state as one would expect from a variational principle. In fact,
for $N=2$ and $\nu=1$ Eq.~\re{ncnh10} yields a value below the exact
ground state (see Fig.~1). This is a consequence of the fitting of
$\delta$ to the result of Wu Ref.~\ci{wu}, which for $N=2$ leads to a
square root dependence in $\nu$, while the exact result for this case
gives a linear dependence. On the other hand, Eq.~\re{ncnh10} applies
for the whole range of parameters of the system $N, \nu$, and
$\omega_0$.

\begin{figure}
\begin{center}
\includegraphics[angle=0,width=10.5cm,scale=1.0]{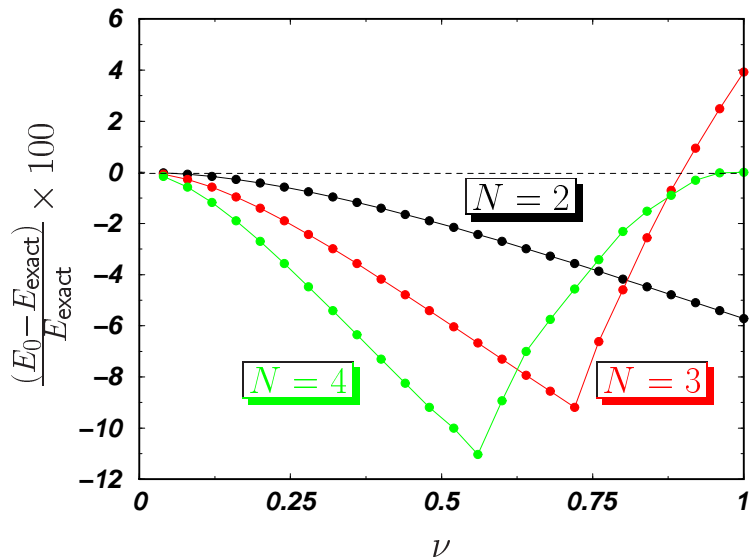}
\caption{
Relative deviation (in percent) of the approximate ground state energy
$E_0$, Eq. \protect \re{ncnh10}, from the exact ground state energy
for 2,3 and 4 anyons in a parabolic confinement potential.
}
\end{center}
\label{fig2}
\end{figure}

\section{Non-interacting case and $H \ne 0$}

In this Section we include an external homogeneous magnetic field. In
the presence of an external magnetic field $\vec H$ the statistical
factor $\nu$ may change sign because $\nu=e\phi/2\pi \hbar$ is a
fraction of the flux quantum carried by each anyon, $\phi_0=2\pi \hbar
c /|e|$, and this flux can be antiparallel to the magnetic field
\ci{ler,vo}. The Hamiltonian is invariant under the transformation
$(x_k,y_k,\nu,\beta)\rightarrow (x_k,-y_k,-\nu,-\beta)$, where
$\beta=eH/|eH|$, and thus the energy spectrum is invariant under
$(\nu,\beta)\rightarrow (-\nu,-\beta)$ (see Ref. \ci{vo}). The spectrum
only depends on $|\nu|$, $\nu \beta$ and the cyclotron frequency
$\omega_c$ (apart from $N$ and $\omega_0$).

The real part of the local energy is given by
\be
\ba{l}
Re E_L(\vec R)=\ds\sum_{k=1}^N \left [\alpha+\dsf{x_k^2+y_k^2}{2}
\left (1-{\alpha}^2 +\dsf{\omega _c^2}{4\omega _0^2} \right )+
\dsf{|\nu|^2}{2}(\vec A_{\nu}(\vec r_k))^2+\right.\\
\qquad\qquad\qquad\qquad\left.
+\dsf{\nu \beta \omega _c}{2 \omega _0} \ds\sum_{j\not=k}^N
\dsf{\vec{r}_{kj}\cdot\vec{r}_k}{|\vec r_{kj}|^2}\right].
\lab{niyh1}
\ea
\ee
We need to compute the contribution coming from the last term in $Re
E_L$. To this end, we have to solve the integral $\int \Psi_T(\vec R)\
\dsf{\vec r_{kj} \cdot \vec r_k} {|\vec r_{kj}|^2}\Psi_T(\vec R) \ d\vec
R$ ($N(N-1)$ integrals of this form contribute to the energy). With the
help of the integral \ci{as}
\be
\int_{0}^{\infty} e^{-a^2 x^2}I_{\nu}(bx) \ dx=\dsf{\pi^{1/2}}{2a}
e^{\frac{b^2}{8a^2}} I_{\frac{\nu}{2}}\left(\dsf{b^2}{8a^2}\right) \ ,
\lab{niyh2}
\ee
where $Re \ \nu>-1$, $Re \ a^2>0$ ($I_{\nu}(x)$ is the modified Bessel
function), and  \ci{d}
\be
\int_{0}^{\infty} e^{- a x} \sinh(b x) \ dx=\dsf{b}
{a^2-b^2}
\lab{niyh3}
\ee
with $ a > b \geq 0$, one gets
\be
\int \Psi_T(\vec R) \ds\sum_{k,j\not=k}^N \dsf{\vec r_{kj}\cdot\vec
r_k} {|\vec r_{kj}|^2} \Psi_T(\vec R) \ d\vec R= \dsf{- N(N-1)}{2}.
\lab{niyh4}
\ee
The averaged real part of the local energy is
\be
E=\frac{N}{2}\left( {\cal N} \ \alpha + \frac{1}{\alpha} \left
(1+\dsf{\omega _c^2 } {4\omega _0^2}\right )\right)  - \dsf{\nu \beta
\omega _c}{2 \omega _0}\dsf{N(N-1)}{2}\ ,
\lab{niyh5}
\ee
and takes its minimum value for
\be
\alpha_0= \left (1+\dsf{\omega _c^2 }{4\omega _0^2}\right )^{1/2}{\cal
N}^{-1/2} \ .
\lab{niyh7}
\ee
The resulting energy minimum is given by (for now we return to standard
units of energy and length)
\be
E_0=N\hbar \left (\omega_0^2 +\dsf{\omega _c^2 } {4}\right )^{1/2}{\cal
N}^{1/2} - \dsf{\nu \beta \hbar \omega _c}{4}N(N-1).
\lab{niyh8}
\ee

As in Section III this expression diverges logarithmically in the limit
of a vanishing cut-off parameter $\delta$. Following the line of
arguments of the previous section, the cut-off parameter - representing
the squared minimum particle distance - should not be zero except for
the bosonic limit $\nu=0$. Having this in mind we determine $\delta$ by
fitting to known exact results for the ground state energy.

To establish these results we calculate the fermion ground state energy
from the single-particle spectrum of the 2$D$ harmonic oscillator in an
external magnetic field perpendicular to the 2$D$ plane (Fock-Darwin
spectrum \ci{fd} ) (see also \ci{gal})
\be
E_{n,m}=\hbar \sqrt{\omega_0^2+ \dsf{ \omega_c^2}{4}} \ (2n+|m|+1)+
m\dsf{\hbar \omega_c}{2} \ .
\lab{niyh9}
\ee
In Eq.~\re{niyh9} $n$ and $m$ are the radial and angular momentum
quantum numbers, respectively. The ground state energy of $N$ spinless
fermionic particles is the sum of the $N$ lowest single-particle
energies (Pauli exclusion principle). Following \ci{kh} we introduce a
parameter $z=R/P$, where $R=\hbar \omega _c/2$ and $P=\hbar (\omega_0^2
+\omega _c^2/4)^{1/2}$ and express the ground state energy in units of
$P$. The parameter $z$ changes between 0 and 1 when the external
magnetic field is changed between 0 and infinity. The Fock-Darwin
spectrum is characterized by level crossings. These crossings,
occurring at $z=z_b,z_2,z_3,...z_l$, have to be considered in
evaluating the ground state energy, their number therefore depending
upon $N$. Every interval between level crossings is characterized by
its own expression for the ground state energy.  However, only for the
intervals $0\leq z \leq z_b$ and  $z_l\leq z \leq 1$  one can write
down the expressions for the energy as a function of the number of
particles $N$. For the sake of clarity let us consider some special
values  of $N$. The cases with one and two fermions are not affected by
crossings. For $N=3$ we have one crossing at $z_b=z_l=1/3$ and two
expressions for the ground state energy: $E/P=5$ and $E/P=6-3z$. This
crossing point coincides with the one considered in Refs. \ci{kh,kmo}
for three anyons. The case with $N=5$ has two crossings $z_b=1/3$ and
$z_l=3/5$ and three expressions for energy: $E/P=11-2z$, $E/P=12-5z$
and $E/P=15-10z$ in the intervals $0 \leq z \leq z_b$, $z_b\leq z \leq
z_l$ and $z_l\leq z \leq 1$, respectively. On the basis of these
special cases one can make the following generalizations:

\begin{itemize}
\item
There are $N_{cl}-K$ crossing points $z=z_b,z_2,z_3,...z_l$ for
$N_{cl}$ fermions in $K$ closed shells. Therefore, there are
$N_{cl}-K+1$ expressions for the ground state energy of fermions and
ground state spectra of anyons for this number of particles.

\item
One can write the expression $z_b=1/(2K-1)$ for $K$ closed shells. The
last crossing point does not depend on $K$, it is a function of $N$:
$z_l=(N-2)/N$ and, thus, it is applicable for any number $N$.

\item
One can write the expression for the ground state energy in the
interval $0 \leq z \leq z_b$ (we choose the smallest $z_b$ for all
particles filling the given shell) or $0\leq \omega _c \leq \omega_0
/(K(K-1))^{1/2}$ for $K\geq2$.
\be
E/P\approx N^{3/2}+zS \ .
\lab{eappr}
\ee
This expression is approximate (to within $6\%$ accuracy (see Section
III)) for $N\not=N_{cl}$, where $S=\ds\sum_{j=0}^{N-N_l-1}(-N_s+2j)$,
$N_s$ is the integer part of $[-1+(1+8N)^{1/2}]/2$ and
$N_l=N_s(N_s+1)/2$, and becomes exact in the form
$E/P=N_{cl}\sqrt{(1+8N_{cl})}/3$ for closed shells, i.e., $N=N_{cl}$.


\item
The expression for the ground state energy in the interval $z_l \leq z
\leq 1$ or $\omega _c \geq \omega_0 (N-2)/(N-1)^{1/2}$, determined by
the lowest levels from each shell in agreement with the Fock-Darwin
formula Eq.~\re{niyh9}, is
\be
E/P=N[(N+1)-z(N-1)]/2.
\lab{enn}
\ee
\end{itemize}

Having discussed the case of fermions ($|\nu|=1$) we now determine the
approximate expressions of the ground state energy of $N$
non-interacting anyons for the two interesting ranges of weak $0\leq
\omega _c \leq \omega_0 /(K(K-1))^{1/2}$ and strong $\omega _c \geq
\omega_0 (N-2)/(N-1)^{1/2}$ magnetic fields. In the weak
magnetic field regime it is
\be
E_0=P N_{cl}[1+|\nu|(N_{cl}-1)]^{1/2}
\lab{cls}
\ee
for closed shells and
\be
E_0= P N[1+|\nu|(N-1)]^{1/2}
+\nu \beta RS
\lab{ncls}
\ee
otherwise. Note that these expressions coincide with Eq. \re{ncnh10}
in the absence of an external magnetic field, i.e., $\omega _c=0$.
We find the cut-off parameters by equating these expressions to Eq.
\re{niyh8}. The result for closed shells is
\be
\delta= \dsf{1}{2} \exp \left[ -\dsf{1}{|\nu|}\left(1+|\nu|[(N_{cl}-1)
(G+z^2/4)-G] + \dsf{\nu \beta z}{|\nu|}[1+|\nu|(N_{cl}-1)]^{1/2}\right)
\right]r_0^2
\lab{dlt1}
\ee
while for open shells
\be
\delta= \dsf{1}{2} \exp \left[ -\dsf{1}{|\nu|}\left(1+|\nu| G(N-2)+
\dsf{2\nu \beta zT}{|\nu|(N-1)}
[1+|\nu|(N-1)]^{1/2}+\dsf{|\nu|z^2T^2}{(N-1)} \right)\right]r_0^2 \ ,
\lab{dlt2}
\ee
where $N\geq2$ and $T=(N-1)/2+S/N$.

In the high magnetic field regime (or for weak confinement) the
Fock-Darwin single-particle energies tend toward Landau levels, the
lowest energy state having the quantum numbers $n=0$ and $m\leq 0$ and
energy $R$. In this limit the ground state energy of $N$ particles is
$NR$, independent of $|\nu|$ \ci{spor2,il,kh}. This exact result can
be reproduced with
\be
{\cal N}^{1/2}=1+ \nu\beta(N-1)/2
\lab{niyh10}
\ee
which gives the cut-off parameter
\be
\delta= \dsf{1}{2} \exp \left[ -\dsf{1}{4|\nu|}[4\nu \beta /|\nu|+
|\nu|(N(4G+1)-(8G+1))] \right]r_0^2.
\lab{dlt3}
\ee
Using this choice in the general formula Eq.~\re{niyh8} we arrive at
the closed analytic expression for the approximate ground state
energy
\be
E_0=PN[1+ \nu \beta (N-1)/2] - \nu \beta RN(N-1)/2.
\lab{niyh11}
\ee

Besides the high magnetic field (or weak confinement) limit used here
to fix the cut-off parameter $\delta$, the expression obtained for
$E_0$ reproduces the exact ground state energy of non-interacting
fermions (the sum of the $N$ lowest Fock-Darwin energies) in the whole
magnetic field range beyond the last crossing of the $N$th level with
$n=0, m=-(N-1)$, which defines the so-called {\it maximum density
droplet} \ci{hawrylak}. Finally, approximate expressions Eqs. \re{cls},
\re{ncls} and \re{niyh11} give the ground state energy $E_0=P N$ of $N$
bosons ($\nu=0$) in a magnetic field and harmonic confining potential.

The separate discussion provided here for small and large magnetic
fields is in correspondence with the treatment of the $N=3$ case
discussed in Refs.~\ci{kmo,kh}.

\section{Coulomb-interacting case and $H=0$}

We now include the effect of the Coulomb repulsions between anyons
$\dsf{r_0}{2a_B}\ds\sum_{k,j\not=k}^N\dsf{1}{|\vec r_{kj}|}$ in the
expression for the real part of local energy $Re E_L(\vec R)$, Eq.
\re{ncnh1}, but in a vanishing external magnetic field. Here, as in
Section III, we assume that $\nu\equiv |\nu|$.

The Coulomb interaction part contributes with $N(N-1)$ integrals of the
form $\int \Psi_T(\vec R)\ \dsf{1}{|\vec r_{kj}|}\Psi_T(\vec R) \ d\vec
R$. These integrals can be evaluated using Eq. \re{niyh2} and \ci{gr}
\be
\int_{0}^{\infty} e^{-a x}I_{\nu}(b x) \ dx=\dsf{b^{\nu}}
{\sqrt{a^2-b^2} \ (a+\sqrt{a^2-b^2})^{\nu}},
\lab{13}
\ee
where $Re \ \nu>-1$ and $Re \ a>|Re \ b|$. The result is
\be
\int \Psi_T(\vec R) \dsf{1}{|\vec r_{ij}|}\Psi_T(\vec R) \ d\vec R=
\left(\dsf{\pi\alpha}{2}\right)^{1/2} \ .
\lab{14}
\ee
The averaged (real part of the) local energy is
\be
E=\frac{N}{2}\left( {\cal N} \ \alpha + \frac{1}{\alpha}+2{\cal M}
\ \alpha^{1/2}\right)\ ,
\lab{15}
\ee
with
\bea
\ba{l}
{\cal M}=\left(\dsf{\pi}{2}\right)^{1/2}
\dsf{N-1}{2}\dsf{r_0}{a_B}.
\lab{16}
\ea
\eea
The extremum condition $\dsf{dE}{d\alpha}=0$ leads to the equation
\be
X^4-{\cal M}X-{\cal N}=0
\lab{17}
\ee
for $X=1/\alpha^{1/2}$. Two complex and two real solutions of this
equation can be found by the Descartes-Euler method \ci{kk}. The
minimum energy is given by the expression
\be
E_0=\dsf{N}{2}\left[\dsf{\cal N}{X_0^2}+X_0^2+\dsf{2{\cal M}}{X_0} \right]
\lab{18}
\ee
and it is achieved at the point
\be
X_0=(A+B)^{1/2}+[-(A+B)+2(A^2-AB+B^2)^{1/2}]^{1/2} \ ,
\lab{19}
\ee
where
\bea
\ba{l}
A=\left[{\cal M}^2/128+\left(({\cal N}/12)^3+({\cal M}^2/128)^2
\right)^{1/2} \right]^{1/3},\\
B=\left[{\cal M}^2/128-\left(({\cal N}/12)^3+({\cal M}^2/128)^2
\right)^{1/2} \right]^{1/3}.
\lab{20}
\ea
\eea
Again, the ground state energy $E_0$, Eq. \re{18}, has a logarithmic
divergence in the limit $\delta \rightarrow 0$. Assuming that $\cal N$
can be regularized, one can recognize two limits of interest. One
corresponding to weak correlations, $r_0/a_B\ll1$,
\be
E_0 \approx N \left({\cal N}^{1/2}+\dsf{\cal M}{{\cal N}^{1/4}}\right)
\lab{21}
\ee
and another for strong correlations, $r_0/a_B\gg1$,
\be
E_0\approx\dsf{3N}{2} \left({\cal M}^{2/3}+\dsf{\cal N}{3{\cal
M}^{2/3}}\right).
\lab{22}
\ee
In order to determine the cut-off parameter $\delta$, and due to the
lack of analytic results, we need to fit to known numerical results for
the ground state energy at special values of the parameter $r_0/a_B$.

\begin{figure}
\begin{center}
\includegraphics[angle=0,width=10.5cm,scale=1.0]{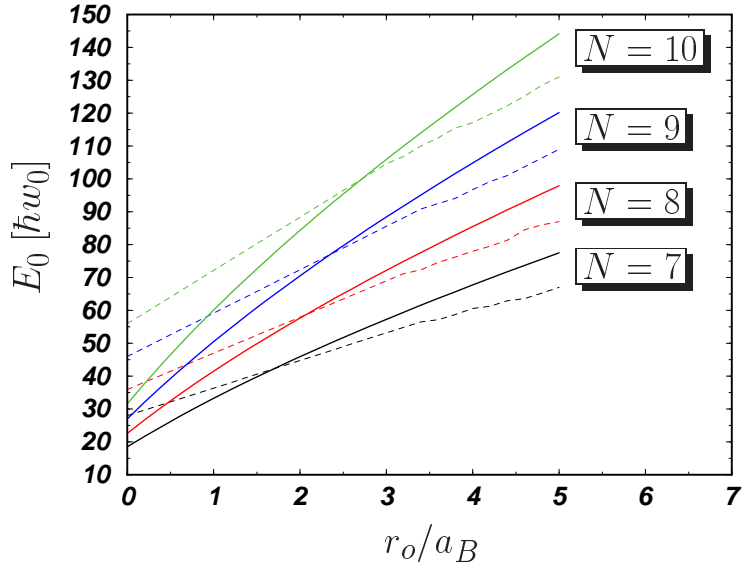}
\end{center}
\caption{
Coulomb interaction parameter $r_0/a_B$ dependence of the ground
state energy for 7 -- 10 electrons calculated by variational
\protect\ci{b1} and  fixed-node quantum Monte Carlo methods
\protect\ci{b3} - dashed curves (results of both calculations are
indistinguishable in these curves) and by formula \protect\re{18} -
solid curves.}
\lab{fig3}
\end{figure}

In Fig.~3 we compare the ground state energies calculated for 7-10
electrons using Eq. \re{18}, with the {\it non-interacting}  ${\cal
N}=1+\nu(N-1)$, to variational \ci{b1} and fixed-node quantum Monte
Carlo calculations \ci{b3} (see also \ci{harju}).

From Eqs. \re{21} and \re{22} follows that the contribution of the
statistics (dependence upon $\nu$) in the ground state energy is
important for weak, and negligible for strong, Coulomb correlations
$r_0/a_B$. For large values of $r_0/a_B$ one can compare the dependence
of the ground state energy, Eq. \re{22}, with the estimate given in
Ref. \ci{yang}. The asymptotic behavior of the ground state energy with
$N$ derived from our expression is $E_0\sim N^{5/3}$ (as in Ref.
\ci{yang}).

\section{Coulomb-interacting case and $H \ne 0$}

Finally, we consider the case of a confined Coulomb anyon gas in an
external magnetic field. The resulting expression for the averaged
local energy is
\be
E=\frac{N}{2}\left( {\cal N} \ \alpha + \frac{1}{\alpha}\left
(1+\dsf{\omega _c^2 } {4\omega _0^2}\right )+2{\cal M} \ \alpha^{1/2}
\right)  - \dsf{\nu \beta \omega _c}{2 \omega _0}\dsf{N(N-1)}{2}\ .
\lab{a16}
\ee
Following steps similar to previous sections we obtain the ground state
energy (in standard units)
\be
E_0=\frac{N}{2}\hbar\omega_0 \left[\dsf{\cal
N}{\bar{X}_0^2}+\bar{X}_0^2 \left(1+\dsf{\omega _c^2 } {4\omega
_0^2}\right ) - \dsf{\nu \beta \omega _c}{2 \omega _0} \
(N-1)+\dsf{2{\cal M}}{\bar{X}_0} \right],
\lab{a19}
\ee
where $\bar{X}_0$ is formally the same expression as Eq. \re{19} after
replacing ${\cal N} \rightarrow {\cal N}_1={\cal N}/(1+\omega _c^2/(4
\omega _0^2))$ and ${\cal M} \rightarrow {\cal M}_1={\cal M}/(1+\omega
_c^2/(4 \omega _0^2))$ in Eq. \re{20}. The
asymptotic expressions are
\be
E_0 \approx N\hbar\omega_0 \left (\left ({\cal N}_1^{1/2}+\dsf{{\cal
M}_1}{{\cal N}_1^{1/4}} \right)\left(1+\dsf{\omega _c^2} {4\omega
_0^2}\right ) - \dsf{\nu \beta \omega _c}{2 \omega
_0}\dsf{(N-1)}{2}\right )
\lab{a20}
\ee
and
\be
E_0\approx\dsf{3N\hbar\omega_0}{2} \left (\left  ({\cal
M}_1^{2/3}+\dsf{{\cal N}_1}{3{\cal M}_1^{2/3}}
\right)\left(1+\dsf{\omega _c^2} {4\omega _0^2}\right ) - \dsf{\nu
\beta \omega _c}{3\omega_0}\dsf{(N-1)}{2}\right ).
\lab{a21}
\ee
for very small and very large values of $r_0/a_B$, respectively.  We
regularize the logarithmic divergence by fitting $\delta$ to known
numerical results for the ground state energy of quantum dots in
external magnetic fields.

\begin{figure}
\begin{center}
\includegraphics[angle=0,width=10.5cm,scale=1.0]{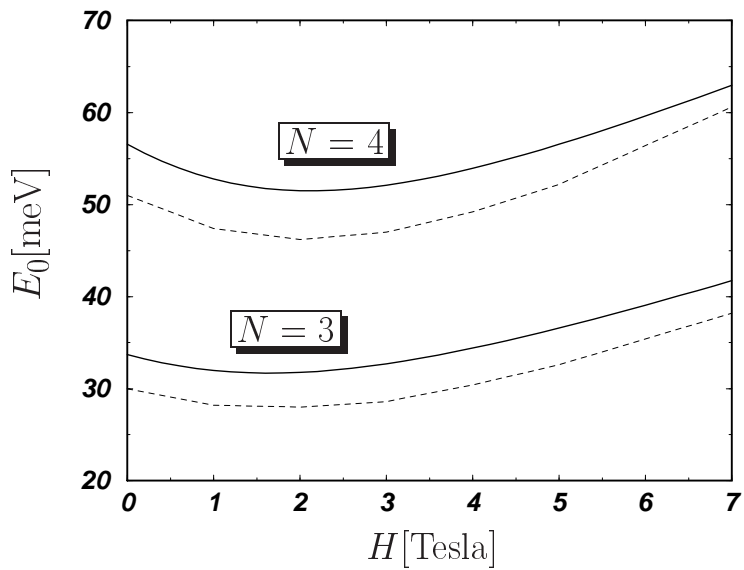}
\end{center}
\caption{
Magnetic field, $H$, dependence of the ground state energy for
$N=3$ and $N=4$ spin-polarized electrons in a harmonic potential
calculated in Ref. \protect \ci{vnus} (the dashed curves), and using
Eq. \protect \re{a19} (the solid curves). As in Ref. \protect \ci{vnus}
we used $\hbar \omega_0=3.37$ meV  ($r_0/a_B=\sqrt{H^*/(\hbar
\omega_0)}$, where the effective Hartree $H^*$ is equal to
$H^*\simeq11.86$ meV).
}
\lab{fig4}
\end{figure}

We compare our results with the ground state calculations of Ref.
\ci{vnus} for $GaAs$ dots in an external magnetic field. The results of
these calculations are very close to the results of Ref. \ci{haw} and
Ref. \ci{b2} computed by exact diagonalization and quantum Monte Carlo
methods, respectively. For $GaAs$ quantum dots $M^*=0.067M$, and the
dielectric constant is $\epsilon=12.4$. Therefore, the effective Bohr
radius $a^*_B=\hbar^2\epsilon/M^*e^2$ is $a^*_B\simeq97.90$\AA and the
unit of energy -- the effective Hartree ($H^*=M^*e^4/(\epsilon^2
\hbar^2)$) is $H^*\simeq11.86$ meV. The cyclotron frequency is
$\omega_c=eH/(M^*c)$ (for simplicity, in this part of our work we
assume $\nu\equiv  |\nu|$, $e\equiv |e|$ and $H\equiv|H|$ while a
correct combination of the signs of these quantities is given in
Eqs.~\re{a19} - \re{a21}). Thus, the energy quanta for this frequency
is $\hbar \omega_c=1.7269\cdot H\cdot$ (meV/$T$). Here we took into
account that $\hbar \omega_c=\hbar e H/(M^*c)=2M\mu_B H/M^*$, the Bohr
magneton is $\mu_B=e\hbar /(2Mc)=0.05785$ meV/$T$ and the magnetic
field $H$ is measured in Tesla ($T$) magnetic units. The Coulomb
interaction parameter $r_0/a_B$ in our case is equal to
$r_0/a_B=\sqrt{H^*/(\hbar \omega_0)}$.

To compare with the results of Ref. \ci{vnus} for spin polarized
electrons, we calculated the ground state energy using Eq. \re{a19}
with the expression ${\cal N}=N$ (i.e., {\it non-interacting} ${\cal
N}$ with $\nu=1$), for three and four particles as a function of the
magnetic field strength $H$. As in Ref. \ci{vnus} we considered $\hbar
\omega_0=3.37$ meV. Comparison of these results is displayed in Fig. 4.
The deviation of our results with respect to the ones given in Ref.
\ci{vnus} is no more than $10 \%$.

It turns out that this expression for $\cal N$  with $\nu=1$ is
appropriate for the description of a small number of electrons
($N=3,4,5,6$) and not large magnetic fields. For large number of
particles and a wide range of magnetic fields one can write the
approximate expression for $\cal N$ (here we return to the original
signs of $\nu$, $e$ and $H$)
\be
{\cal N}=F \left[(1+ |\nu|(N-1))^{1/2}+\dsf{\nu \beta \omega _c
(N-1)}{4 \omega _0}- \left(\dsf{\nu \beta \omega _c N^{1/2}}{\omega
_0}\right)^{1/2}\right]^2
\lab{a22}
\ee
and thus for $\delta$
\be
\delta= \dsf{1}{2} \exp\left[ -\dsf{F}{|\nu|} \left(1+|\nu| G(N-2)+
\dsf{|\nu|\omega _c^2 D}{4 \omega _0^2}+\dsf{\nu \beta \omega _c C}
{2|\nu| \omega _0}-Q \right)
\right]r_0^2
\lab{a23}
\ee
with
\be
F=\dsf{1}{1+\dsf{|\nu|^2 \omega _c^2}{4 \omega _0^2}},
\lab{a24}
\ee

\be
D=\dsf{N^2-2N-3}{4(N-1)}+|\nu|^2 G (N-2),
\lab{a25}
\ee
\be
C=\dsf{2N^{1/2}}{N-1}+(1+|\nu|(N-1))^{1/2}-
\left(\dsf{\nu \beta \omega _c N^{1/2}}{\omega _0}\right)^{1/2} ,
\lab{a26}
\ee
and
\be
Q=2\left[\dsf{\nu \beta \omega _c N^{1/2}}{|\nu|^2 \omega _0 (N-1)^2}
(1+|\nu|(N-1))\right]^{1/2} \ ,
\lab{a27}
\ee
for $N\geq 2$. Here we took into account that $\cal N$ depends weakly
on the Coulomb parameter $r_0/a_B$ (the results indicated in Fig. 3
have been calculated with $\cal N$ not having this parameter dependence).

In Fig. 5 we compare the ground state energy calculated with the
expression Eq. \re{a19}, (using Eq. \re{a22} for $\cal N$ with
$|\nu|=1$)  for $16\leq N\leq 40$ and $0\leq \omega _c/\omega _0\leq
20$ ($r_0/a_B=1.911$), with the calculations for a classical system of
electrons of Ref. \ci{kainz}. The deviation is maximal (no more than
$15 \%$) in the range $1/(K(K-1))^{1/2}\leq \omega_c/\omega _0\leq
(N-2)/(N-1)^{1/2}$, where $K$ is the number of closed shells. This
range of magnetic fields corresponds to the crossings of Fock-Darwin
levels (see Ref. \ci{fd} and  Section IV) and, therefore, the single
particle ground state energy of electrons changes many times as $\omega
_c$ increases. One can suppose that the Coulomb interaction shifts the
levels but the qualitative structure of the many particle ground state
is still complex. Thus, in this range of parameter $\omega_c/\omega_0$
the expression for $\cal N$ is not uniquely defined. We could not find
a more appropriate expression for $\cal N$ than Eq. \re{a22}, for the
magnetic fields indicated in Fig. 5.

\begin{figure}
\begin{center}
\includegraphics[angle=0,width=10.5cm,scale=1.0]{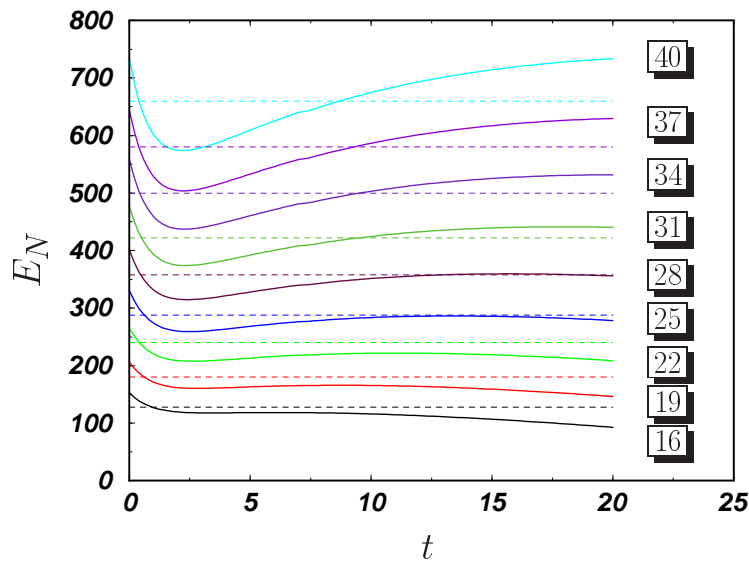}
\end{center}
\caption{
Ground state energy $E_N=(E_0-N \hbar \omega)/(\hbar \omega_0)$ for 16
- 40 electrons calculated using the expression Eq. \protect \re{a19}
for $r_0/a_B=1.911$, applying the expression for $\cal N$  Eq. \protect
\re{a22} with $|\nu|=1$ - solid curves, and energy for classical
electrons (Ref. \protect \ci{kainz}) - dashed lines. Here
$\omega=(\omega^2_0+\omega_c^2/4)^{1/2}$ and $t=\omega_c /\omega _0$.
}
\lab{fig5}
\end{figure}

\section{Conclusion}

We have used the anyon concept combined with a variational calculation
to obtain an analytic closed-form expression for the approximate ground
state energy of $N$ non-interacting and Coulomb-interacting particles
in a 2$D$ harmonic confining potential, with and without an external
magnetic field. The crucial point of this approach is the appearance of
a logarithmic divergence connected with a cut-off parameter, when
evaluating the contribution of the gauge field vector potential.
Following arguments from the literature, according to which the cut-off
parameter cannot be zero (except for the bosonic limit), we used it to
fit our results to exact and numerical ground state energies known for
special values of the system parameters. In doing so we provided closed
analytic expressions for the approximate ground state energy depending
upon $|\nu|, N, \omega_0, r_0/a_B$ and $\omega_c/\omega_0$.

\section{Acknowledgments}
One of the authors (B.A.) acknowledges support from the DAAD
organization of the Government of Germany. He also thanks the Los
Alamos National Laboratory and RIISE, Hiroshima University of Japan,
for their kind hospitality and financial support.

\end{document}